\documentclass[reprint,aps,numerical,superscriptaddress,floatfix,nofootinbib]{revtex4-1}
\usepackage{amssymb}
\usepackage{amsthm}
\usepackage[figuresright]{rotating}
\usepackage{graphicx}
\usepackage{tikz}			% erlaubt mit Latex zu Zeichnen
\usetikzlibrary{decorations.pathmorphing}       % for snake lines
\usetikzlibrary{decorations.pathreplacing}      % for braces lines
\usetikzlibrary{positioning}                    % for more precise positioning
\usetikzlibrary{arrows}                         % for arrow heads
\usetikzlibrary{patterns}                       % for fill patterns
\usetikzlibrary{shapes}                         % for special node shapes
\usetikzlibrary{scopes}
\usepackage{textcomp}
\usepackage[T1]{fontenc}
\usepackage[tight]{units}
\newcommand{\figref}[1]{figure~\ref{#1}}

\usepackage{hyperref}
\usepackage{lineno}
\begin{document}
\title{The COSINUS project - perspectives of a  NaI scintillating calorimeter for dark matter search}
%%%%%%%%%%%%%%%%%%%%%%%%%%%%%%%%%%%%%%%%%%%%%%%%%%%%%%%%%%%%%%%%%%%%%

%%%%%%%%%%%%%%%%%%%%%%%%%%%%%%%%%%%%%%%%%%%%%%%%%%%%%%%%%%%%%%%%%%%%%
\newcommand{\LNGS}{\affiliation{INFN - Laboratori Nazionali del Gran Sasso, Assergi (AQ) I-67010 - Italy}}
\newcommand{\mpi}{\affiliation{Max-Planck-Institut f\"ur Physik, D-80805 M\"unchen - Germany}}
\newcommand{\AQ}{\affiliation{Dipartimento di Scienze Fisiche e Chimiche - Universit\`{a} degli studi dell'Aquila, I-67100 Coppito (AQ) - Italy}}
\newcommand{\GSSI}{\affiliation{Gran Sasso Science Institute, I-67100 L'Aquila - Italy}}
\newcommand{\INFNMilano}{\affiliation{INFN - Sezione di Milano Bicocca, Milano I-20126 - Italy}}
\newcommand{\Bicocca}{\affiliation{Dipartimento di Fisica, Universit\`{a} di Milano-Bicocca, Milano I-20126 - Italy}}
\newcommand{\Vienna}{\affiliation{Institut f\"ur Hochenergiephysik der \"Osterreichischen Akademie der Wissenschaften, A-1050 Wien - Austria and Atominstitut, Vienna University of Technology, A-1020 Wien - Austria}}
%%%%%%%%%%%%%%%%%%%%%%%%%%%%%%%%%%%%%%%%%%%%%%%%%%%%%%%%%%%%%%%%%%%
\author{G.~Angloher}
  \mpi

\author{D.~Hauff}
 \mpi

\author{L.~Gironi}
\INFNMilano
\Bicocca

\author{C.~Gotti}
\INFNMilano

\author{G.~Pessina}
\INFNMilano

\author{A.~G\"utlein}
\Vienna

\author{M.~Maino}
\INFNMilano

\author{S.S.~Nagorny}
 \GSSI

\author{L.~Pagnanini}
\GSSI

\author{F.~Petricca}
 \mpi

\author{S.~Pirro}
 \LNGS

\author{F.~Pr\"obst}
 \mpi
 
\author{F.~Reindl}
\email[corresponding author:]{florian.reindl@mpp.mpg.de}
\mpi

\author{K.~Sch\"affner}
\email[corresponding author:]{karoline.schaeffner@lngs.infn.it}
\GSSI

 \author{J.~Schieck}
 \Vienna
 
\author{W.~Seidel}
 \mpi
 
%%%%%%%%%%%%%%%%%%%%%%%%%%%%%%%%%%%%%%%%%%%%%%%%%%%%%%%%%%%%%%%%%%%%
\begin{abstract}
The R\&D project COSINUS (Cryogenic Observatory for SIgnatures seen in
Next-generation Underground Searches) aims to develop a cryogenic
scintillating calorimeter using NaI as target crystal for direct dark
matter search. Dark matter particles interacting with the detector
material generate both a phonon signal and scintillation light. While the
phonon signal provides a precise determination of the deposited energy, the simultaneously measured scintillation light allows for a particle identification on an event-by-event basis, a powerful tool to study material-dependent interactions, and to suppress backgrounds. Using the same target material as the DAMA/LIBRA collaboration, the COSINUS technique may offer a unique possibility to investigate and contribute information to the presently controversial situation in the dark matter sector. We report on the dedicated design planned for the NaI proof-of-principle detector and the objectives of using this detection technique in the light of direct dark matter detection.
\end{abstract}
\maketitle
\section{Introduction}
\label{intro}
In the era of precision cosmology we know that dark matter \cite{planck_collaboration_planck_2015} is five times more prevalent than baryonic matter in the Universe, and experimental evidence of dark matter so far solely relies on gravitational interaction.\\
Among a long list of hypothetical new particles WIMPs (Weakly Interacting Massive Particles) provide compelling arguments as this class of candidates, with a mass in the (GeV-TeV)/$c^{2}$ region and weak-scale interactions, may provide for a relic density that matches observation. Participation of such particles in weak-scale interactions would also allow for a direct detection in earth-bound detectors via the process of elastic scattering off atomic nuclei \cite{Goodman}.\\
Since the expected recoil energies are in the sub-keV to keV regime, depending on the mass of the dark matter particle and the target material, the challenge of such direct searches is to combine an ultra-low background with a highly sensitive detection apparatus.\par
Present-date, the field of direct dark matter search is very active, with numerous experiments all over the world aiming for further increase in size and sensitivity within the next years.\\
Direct searches are designed to register signals in the detector induced by the interacting particles. The involved detection channels induced by such scattering processes include light, charge and/or phonon signals. Experiments are either based on a single or on a dual channel readout.  Which of the three channels are taken in consideration depend on the respective target and the applied detection technique. From the first channel the energy of the interacting particle is determined. In most cases the second channel serves to identify the nature of the interacting particle, thereby providing a powerful tool to discriminate dark matter signals from background.\\
Dark matter searches may be further divided with respect to the approach they use for detection: A numerous class of experiments is optimized to look for nuclear recoil events in a certain energy window where dark matter particles can induce signals. The number of observed recoil events is consequently combined with model assumptions, concerning e.g.~the density of dark matter, in order to derive properties of dark matter particles, as e.g.~their mass and scattering cross-section.  Somewhat different therefrom, dark matter may also be identified via an annual modulation signal caused by the seasonal variation  of the Earth's velocity with respect to the dark matter halo \cite{drukier_detecting_1986}.\\
At present the situation in the sector of direct dark matter detection is controversial and a definite discovery is still absent. Several experiments \cite{Agnese2013, Cogent, DAMA2013} detect events above the known background level which allow to infer properties of a dark matter particle. In particular, the DAMA/LIBRA collaboration observes in more than 13 annual cycles a statistically robust modulation signal using a total of up to \unit[250]{kg} of radiopure sodium iodide (NaI(Tl)) crystals operated as room temperature scintillating detectors in the Laboratori Nazionali del Gran Sasso (LNGS), a deep underground site in central Italy.\\ 
At the same time, the DAMA/LIBRA signal is not consistent, in the standard elastic scattering scenario, with the null results of other direct detection experiments as listed in legend and caption of  \ref{fig:landscape}. However, in order to compare the results of these experiments with the DAMA/LIBRA modulation signal one has to take into account astrophysical assumptions as well as assumptions on the physics of the scattering process of the dark matter particle with the respective target materials. A pure modulation signal instead can be considered model-independent and, thus, has the potential to give a unique and robust signature of new particles. Nonetheless, future investigations are of pivotal importance to understand the origin/composition of the modulation signal observed by the DAMA/LIBRA collaboration either being due to dark matter or a different, so far not understood, physics phenomenon or background.

\section{Detector Design}
\label{sec:2}
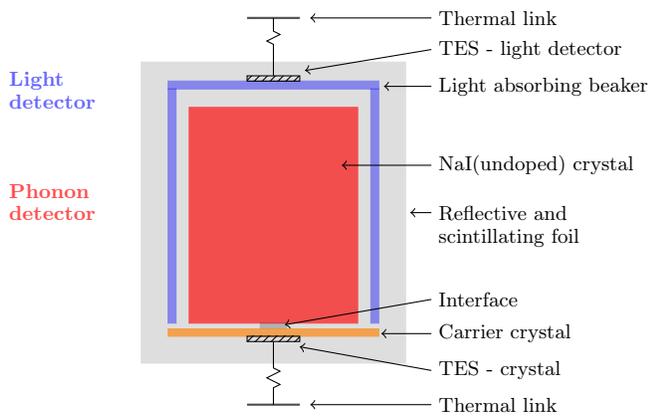
\begin{figure}
\centering
  \begin{tikzpicture}[scale=0.7, every node/.style={scale=0.85}]
   
    \draw[color=black, thick, fill=black, opacity=0.13]
      (-1.5,-2.85) rectangle (3.5,2.85);
    \draw [color=black, thick, opacity=0.7]
      (0.5,3.7) -- (1.5,3.7);
    \draw[decorate, decoration={zigzag, pre length=5, post length=10, segment length=5}]
      (1,3.7) -- (1,2.6);
    \draw[color=blue, fill=blue, opacity=0.4]
      (-1.0,2.5) rectangle (3.0,2.35);
       \draw[color=blue, fill=blue, opacity=0.4]
      (-1.0,2.35) rectangle (-.85,-2.1);
      \draw[color=blue, fill=blue, opacity=0.4]
      (3.0,2.35) rectangle (2.85,-2.1);
    \draw [color=black, pattern=north east lines]
    (0.5,2.6) rectangle (1.5,2.5);
    \draw[color=red, fill=red, opacity=0.65]
      (-.60,2.0) rectangle (2.6,-2.1);
    \draw[color=black, fill=black, opacity=0.2]
    (0.75,-2.1) rectangle (1.25,-2.2);
    \draw[color=orange, fill=orange, opacity=0.65]
      (-1,-2.35) rectangle (3,-2.21);
    \draw[decorate, decoration={zigzag, pre length=10, post length=5, segment length=5}]
      (1,-2.45) -- (1,-3.65);
    \draw[color=black, thick, opacity=0.7]
      (0.5,-3.65) -- (1.5,-3.65);
    \draw[color=black, pattern=north east lines]
    (0.5,-2.35) rectangle (1.5,-2.45);
    \draw[color=black,->] (4.0,3.7) node[color=black, right]{Thermal link} -- (1.7, 3.7);
    \draw[color=black,->] (4.0,3.1) node[color=black, right]{TES - light detector} -- (1.6, 2.7);
    \draw[color=black,->] (4.0,2.4) node[color=black, right]{Light absorbing beaker} -- (3.1, 2.4);
    \draw[color=black,->] (4.0,0.9) node[color=black, right]{NaI(undoped) crystal} -- (2.3, 0.9);
    \draw[color=black,->] (4.0,0.0) node[color=black, right]{Reflective and} -- (3.6, 0.0);
    \draw(4.0,-0.48) node[color=black, text width=2.5cm, right]{scintillating foil};
    \draw[color=black,->](4.0,-3.0) node[color=black, text width=2.3cm, right]{TES - crystal} --(1.5, -2.55);    
    \draw[color=black,->](4.0,-1.65) node[color=black, text width=2.3cm, right]{Interface} --(1.22, -2.13);
    \draw[color=black,->](4.0,-2.3) node[color=black, text width=2.5cm, right]{Carrier crystal} --(3.05, -2.3);
    \draw[color=black,->](4.0,-3.65) node[color=black, text width=2.3cm, right]{Thermal link} -- (1.7, -3.65);
    \draw(-2.5,0.4) node[color=red,thick, opacity=0.7, text width=2.5cm]{\textbf{Phonon}};
    \draw(-2.5,0.0) node[color=red, thick, opacity=0.7, text width=2.5cm]{\textbf{detector}};
    \draw(-2.5,2.5) node[color=blue, thick, opacity=0.6, text width=2.5cm]{\textbf{Light}};
    \draw(-2.5,2.1) node[color=blue, thick, opacity=0.6, text width=2.5cm]{\textbf{detector}};
  \end{tikzpicture}
  \caption{Concept of a COSINUS detector module consisting of an undoped NaI target crystal and a beaker-shaped light detector. Both detectors, operated at milli-Kelvin temperatures, are read out by transition edge sensors (TES) and are surrounded by a reflective and scintillating foil.}
  \label{pic:module_schema}
\end{figure}
The objective of COSINUS (Cryogenic Observatory for SIgnatures seen in Next-generation Underground Searches) is the development of a cryogenic scintillating calorimeter using a NaI crystal as target - thereby providing a NaI detector with the possibility for active particle identification on an event-by-event basis thanks to the dual channel detection approach.\\
Particles scattering in NaI crystals do not only create phonons, but also scintillation photons. At temperatures of few milli-Kelvin, both the phonon signal in the crystal and the signal arising from the absorption of scintillation photons in a suitable light absorber can be detected by the use of superconducting thin film thermometers: As thin film technology is difficult to be directly applied to the hygroscopic NaI crystal, a carrier disk made of a more robust scintillating material (e.g.~CdWO$_{\text4}$) is planned to be used instead. The carrier disk (diameter of \unit[(40-50)]{mm} and about \unit[1]{mm} thickness) is designed to exceed in diameter the size of the NaI crystal and carries the thermometer: A transition edge sensor (TES) of CRESST-type consisting of a thin tungsten film (\unit[200]{nm}, W-TES) directly evaporated onto the carrier crystal. The target crystals of undoped NaI can weigh up to \unit[$\sim$200]{g}, depending on the achievable performance in function of the crystal mass.\\
In order to efficiently reject any $\alpha$-related background, e.g.~a recoiling nucleus, which could mimic a dark matter signal \cite{TUM40alpha}, there must not be any non-active surfaces in the line-of-sight of the target crystal. An elegant way to tag and reject such backgrounds is a completely active surrounding of the crystal. Thus in COSINUS, the light detector is planned to consist of a beaker-shaped work-piece made from high purity silicon serving two purposes: Scintillation light detection and a fully active surrounding of the target crystal. The beaker will exhibit dimensions of about \unit[(40-50)]{mm} in diameter and height, the wall-thickness will be about \unit[600]{$\mu$m}. The front face of the polished silicon beaker will be equipped with a W-TES, optimized in size for the purpose of light detection. Despite being quite macroscopic devices, such kind of light detectors have shown to be highly performing, achieving a baseline noise of below $\sigma\simeq\unit[10]{eV}$\footnote{Two beaker-shaped light detectors were operated in CRESST-II phase 2.} \cite{Becher}. For optimal resolution and because of their low impedance, TESs are usually read out with SQUID amplifiers \cite{Proebst}.\\
With the help of a dedicated copper structure the target crystal will be held inside the beaker-shaped light detector. Since the carrier crystal is chosen to be slightly larger in diameter than the target crystal and the light detector, an almost 100\% coverage of the target crystal should be achievable. A schematic drawing thereof can be seen in \figref{pic:module_schema}.\par 
A particle interaction in the target crystal mainly induces a thermal signal detected by the W-TES. This so-called \textit{phonon-signal} provides a precise measurement of the deposited energy, independent of the type of particle \cite{TUM40alpha, Arnaboldi2010}, often referred to as unquenched channel.\footnote{The measured energy in the crystal is independent of the particle, when the small fraction of the energy, escaping in form of scintillating light, is taken into account.} Interactions which take place in the carrier crystal itself can be discriminated from energy deposits in the NaI crystal by pulse-shape analysis \cite{Becher,CsI_Tech}.\\
Simultaneously to the phonon signal a small fraction of the energy deposited in the NaI crystal is emitted in form of scintillation light and detected by the high-purity silicon light-absorbing beaker. The \textit{light signal} allows to identify the type of interacting particle, as the amount of scintillation light strongly depends on it. Therefore, the light to phonon ratio, referred to as \textit{light yield}, is characteristic for each type of event. Betas and gammas produce the most light and get assigned a light yield value of one by definition. Other types of particles (e.g. $\alpha$-particles, neutrons inducing nuclear recoils off iodine and sodium) exhibit a lower light yield quantified by the so-called \textit{quenching factor} \cite{StraussQF, Tretyak}. The QF is defined as the ratio of scintillation light produced by an interacting particle of type X to the scintillation light produced by a gamma of the same deposited energy.\par
The dual channel detection approach of COSINUS aims to allow for a highly efficient discrimination and rejection of background events ($e^-$/$\gamma$s, $\alpha$-particles) from potential signal events (nuclear recoils) in the NaI target.\\
The low temperatures required for detector operation will be supplied by a $^{\text{3}}$He/$^{\text{4}}$He-dilution refrigerator. The test cryostat of the Max-Planck-Institute is the favored choice for all the measurements necessary for the first basic prototype testing. The hall C CUORE/CUPID R\&D dilution refrigerator, since equipped with passive shielding for means of background reduction, is considered for final prototype testing and a longer performance run. Both refrigerators are located in the LNGS underground site for obvious reasons of shielding against cosmic radiation.\\
The results from CsI measurements give a first reference point on the already achieved performance of an alkali halide crystal operated as cryogenic calorimeter. With a sequence of tests we commissioned the \textit{carrier crystal - absorber crystal concept} and we achieved, with two different CsI crystals, an energy resolution of \unit[0.95]{keV}/\unit[0.70]{keV}, corresponding to a trigger threshold of \unit[4.7]{keV}/ \unit[3.5]{keV} \cite{CsI_Tech}.\footnote{However, a standard CRESST-sized W-TES was used leaving room for future sensitivity improvements via the optimization of the W-TES  for the given application.}

\section{Performance from simulated data} \label{sec:PerformanceSim}
\begin{figure}
\center
\includegraphics[width=0.48\textwidth]{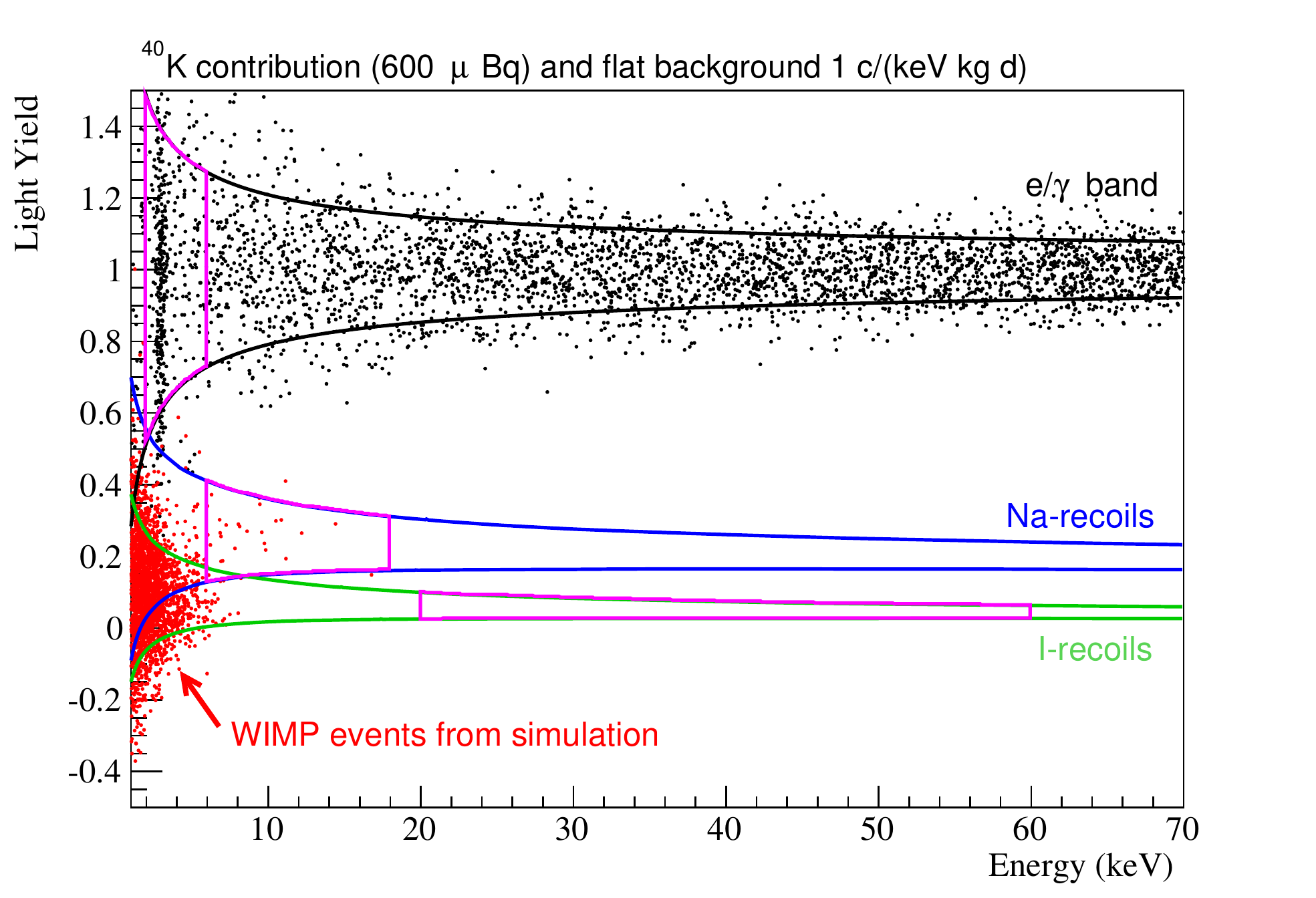}
\caption{Simulated data for an exposure before cuts of \unit[100]{kg-days} applying certain performance assumptions as discussed in detail in the text. The bands correspond to the following event classes: electron recoils in black, scatterings off Na and I in blue and green.  Depicted in black dots are events originating from a flat background contribution of 1 c/(keV kg d) and a $^{\text{40}}$K contamination with an activity of \unit[600]{$\mu$Bq}. The contribution from $^{\text{40}}$K is visible as a line at around \unit[3]{keV} in the e$^{-}$/$\gamma$-band.  A hypothetical WIMP scattering off I and Na nuclei yields the events marked in red. In addition, the magenta boxes indicate the regions which may contribute, apart from the e$^{-}$/$\gamma$-band, to the DAMA/LIBRA energy range of positive modulation detection corresponding to \unit[2-6]{keV$_{ee}$}.}
\label{fig:WIMP_sim}
\end{figure}
We base the performance estimate of a COSINUS detector module on the following assumptions:
\begin{itemize}
  \item light detector energy resolution from standard performance of silicon beaker light detector operated in CRESST-II phase 2 ($\sigma=\unit[0.11]{keV_{ee}}$)
  \item phonon detector resolution of  $\sigma$=\unit[0.2]{keV} (corresponding to a threshold of $5\sigma=\unit[1]{keV}$),
  \item 4\% of deposited energy in the NaI detected in the light detector,
  \item a gross exposure (before cuts) of \unit[100]{kg-days}. Based on results from CRESST-II \cite{CRESST2,CRESST_LISE}, we conservatively estimate an efficiency of \unit[50]{\%} down to an energy of \unit[2]{keV} and a linear decrease to \unit[20]{\%} efficiency at the threshold of \unit[1]{keV},
  \item a flat background contribution down to energy threshold of 1 count / (keV kg d) and an additional contribution from $^{\text{40}}$K at a level of \unit[600]{$\mu$Bq}.
\end{itemize}
In the following the above assumptions will be motivated in more detail.\\
In COSINUS we aim to bring the performance of a NaI-based detector in line with other scintillating bolomoters, rendering an \textit{energy resolution of $\sigma=\unit[0.2]{keV}$}, corresponding to an \textit{energy threshold of \unit[1]{keV}}, feasible. From our understanding the key points to be addressed are the optimization of the sensitivity of the TES itself and of the interface between NaI crystal and carrier crystal.\\
In \cite{CsI_Tech} we measured two CsI-crystals; in average around \unit[7]{\%} of the energy deposited in the crystal were determined as scintillation light (for an e$^{-}$/$\gamma$-event). From measurements in \cite{Nadeau} it is known that CsI roughly emits 3.5 times more energy in form of scintillation light as NaI. However, for COSINUS beaker-shaped light detectors will be used which are found to collect more than twice as much light the light detector of \textit{conventional} CRESST design used for the measurement in \cite{CsI_Tech}. Thus, in total we expect $\unit[7]{\%} / 3.5 \cdot 2 = \unit[4]{\%}$ of the deposited energy in a NaI crystal to be detected in the light detector. \par
Using the above assumptions a simulation is carried out - the outcome is depicted in the light yield-energy plane in \figref{fig:WIMP_sim} with data points arising from simulated contributions of e$^-$/$\gamma$-events in black and a potential signal in red. To stick to the DAMA/LIBRA background budget, an overall flat background contribution of 1 count / (keV kg d) is assumed.\\
NaI crystals typically show a contamination with $^{\text{40}}$K which undergoes an electron capture, either directly to the ground state of $^{\text{40}}$Ar (0.2\% branching ratio) or to an excited state of $^{\text{40}}$Ar de-exciting by the emission of an \unit[1.46]{MeV} gamma (10.55\% branching ratio). Both processes are accompanied by the energy releases of \unit[3.2]{keV} and \unit[3]{keV}, respectively. This contamination makes up for the line at about \unit[3]{keV} in the e$^-$/$\gamma$-band in \figref{fig:WIMP_sim} corresponding to a $^{\text{40}}$K-activity of \unit[600]{$\mu$Bq} as measured by the DAMA/LIBRA collaboration \cite{DAMA_K40}.\\
The bands in \figref{fig:WIMP_sim} are calculated on the basis of the bands determined in the CsI measurement (a detailed description on the underlying model is e.g. given in \cite{StraussQF}), the experience of beaker-shaped light detectors operated in CRESST and the assumptions listed above. Thereby, the lines depict the lower and upper 90\%-boundaries for the bands corresponding to different event classes to be observed in NaI (electron recoils in black, nuclear recoils of Na and I in blue and green, respectively). Thus, in between two corresponding lines 80\% of the events of the respective event class are expected. The widths of the bands is dominated by two effects: Firstly by finite baseline resolution of phonon and light detector and secondly by Poissonian fluctuations in the number of scintillation photons produced. The mean of a band obviously depends on the type of particle and the respective quenching factor. For scatterings off Na and I we use the energy-dependent values reported by Tretyak et al.~\cite{Tretyak}.  \\
In summary, the overall e$^-$/$\gamma$-background (black dots) is given by the $^{\text{40}}$K contamination on top of the constant background level of 1 count / (keV kg d). With the performance and backgrounds assumed, we expect five counts below the mean of the Na-band for a gross exposure of \unit[100]{kg-days} and an energy threshold of \unit[1]{keV}. Thereby, one half of the counts originate from the constant background and the other half from the $^{\text{40}}$K contamination, which roughly corresponds to 0.9\% of the activity in the full double-peak. The expected leakage quickly drops with increasing energy, due to the enhanced separation between e$^-$/$\gamma$-band and Na-band. Additionally, the above numbers show that the leakage is to a large extent caused by the  $^{\text{40}}$K contamination yielding a considerably diminishing leakage for energies above \unit[3]{keV}, where we expect less than one leakage event (for \unit[100]{kg-days} before cuts), which is commonly denoted \textit{background-free}. Considering leakage to the full 80\% Na-band (as depicted in \figref{fig:WIMP_sim}) the values change to 26 events expected above the threshold of \unit[1]{keV}, out of which 13 events are attributed to $^{\text{40}}$K-origin. For the complete Na-band the anticipated leakage drops below one event at \unit[3.9]{keV}.\\
Because of this high discrimination power we can tolerate a moderate level of e$^-$/$\gamma$-background, in particular originating from $^{\text{40}}$K, while still maintaining high sensitivity for a nuclear recoil signal - a distinct feature of COSINUS compared to other NaI-based dark matter searches \cite{ANAIS, DMICE2014, SABRE2015,DAMA2013}.\\
The red events in figure \ref{fig:WIMP_sim} result from a contribution of a hypothetical WIMP of \unit[10]{GeV/$c^{2}$} and \unit[0.0002]{pb} which is consistent with the interpretation of the DAMA/LIBRA modulation signal by Savage et al.~\cite{Savage}. The blue-colored islands in \figref{fig:landscape} correspond to the DAMA/LIBRA signal regions for scatterings of Na and I. For clarity we indicated the assumed dark matter particle mass and nucleon cross-section for the simulated data presented here in form of a blue benchmark point (see figure \ref{fig:WIMP_sim}).\\
 What concerns the distribution of the hypothetical WIMP events we find in the simulation a total of 2386 events in the energy window \unit[(1-6)]{keV}. About 45\% of these events are present in the energy interval from \unit[(1-2)]{keV} and about 53\% within \unit[(2-6)]{keV}. These numbers clearly show the benefit of a low threshold. Despite the conservative assumption on the cut efficiency being \unit[20]{\%} at threshold energy, the number of expected events roughly doubles when lowering the threshold from \unit[2]{keV} to \unit[1]{keV}, which is attributed to the anticipated exponential rise of the dark matter recoil spectrum. Above an energy of \unit[6]{keV} (up to \unit[100]{keV}) we find a total of 46 events.\\
 Since WIMPs are expected to scatter coherently off the nucleus as a whole, the cross-section scales quadratically with the atomic mass number (A$^2$), thus preferring the heavy I over the light Na. As a consequence, the WIMP events in \figref{fig:WIMP_sim} appear almost symmetric to the I recoil band. However, the energy transferred in a scattering for the rather light WIMP considered here is enhanced for light nuclei due to kinematic reasons. Consequently, the majority of the 46 events above \unit[6]{keV} is found in the Na recoil band. For this reason, cryogenic detectors providing low thresholds in combination with light target nuclei currently lead the field of direct dark matter searches in the low mass regime \cite{CRESST_LISE,supercdms_collaboration_wimp-search_2015}.\\
At last we want to discuss and underline the prominent features of such COSINUS detector in contrast to a purely scintillating detector considering the DAMA/LIBRA experiment as representative. In \figref{fig:WIMP_sim} we display magenta colored boxes indicating the different regions that may contribute to the positive modulation signal observed by DAMA/LIBRA in an energy window from \unit[(2-6)]{keV$_{ee}$}. Due to the lack of particle discrimination it remains unknown if the positive modulation signal is made from particles scattering off the electrons or purely off the nuclei in NaI(Tl). \\
 For interactions with electrons the box extends from \unit[(2-6)]{keV} (e$^-$/$\gamma$-band), the same energy range as for DAMA/LIBRA. Instead, interactions on nuclei are quenched in the light channel, hence the energy interval for Na and I recoil events has to be corrected for by the respective quenching factor (energy-dependent QF are taken from \cite{Tretyak}). For Na-recoils (QF$\approx$0.3) the region is restricted to about \unit[(6-20)]{keV} as qualitatively indicated by the magenta box in the Na-recoil band. For I-recoils, due to the even higher light quenching effect (QF$\approx$0.1), the box is confined in an energy region of about \unit[(20-60)]{keV}. \\
As already mentioned, a distinctive feature of the COSINUS technology is the unquenched phonon channel directly measuring the deposited energy quasi independent of the type of interacting particle. Taking into account the aimed for energy threshold of \unit[1]{keV}, this would result in an improvement in detection threshold by a factor of about 6 for Na-recoil events and a factor of about 20 for I-recoils in comparison to the sensitivity demonstrated by DAMA/LIBRA.\\
It deserves mentioning that we already achieved, by operating two scintillating calorimeters based on CsI an energy threshold as low as \unit[4.7]{keV} and \unit[3.5]{keV}, respectively \cite{CsI_Tech}. Thus, such detectors already indicate an increased sensitivity for nuclear recoil events in comparison to the DAMA/LIBRA experiment with further refinements anticipated for the future \cite{CsI_Tech}.

\section{Perspective and Conclusion}
To estimate the sensitivity of the COSINUS detector technology in the standard elastic scattering scenario we simulate 10,000 data sets following the strategy outlined in section \ref{sec:PerformanceSim}. Thereby, we assume e$^-$/$\gamma$-backgrounds only, with a constant background level of 1 count / (keV kg d) and a $^{40}$K-contamination of \unit[600]{$\mu$Bq}. \footnote{Such would acquire a dilution refrigerator in an underground site furnished with adequate passive shielding to significantly reduce amongst others the neutron background. Furthermore, an active muon veto is mandatory to reject any muon-related induced backgrounds in the experimental set-up and the detectors itself.} For each simulated data set a limit on spin-independent WIMP-nucleus scattering is calculated. \\
\begin{figure}
\center
\includegraphics[width=0.49\textwidth]{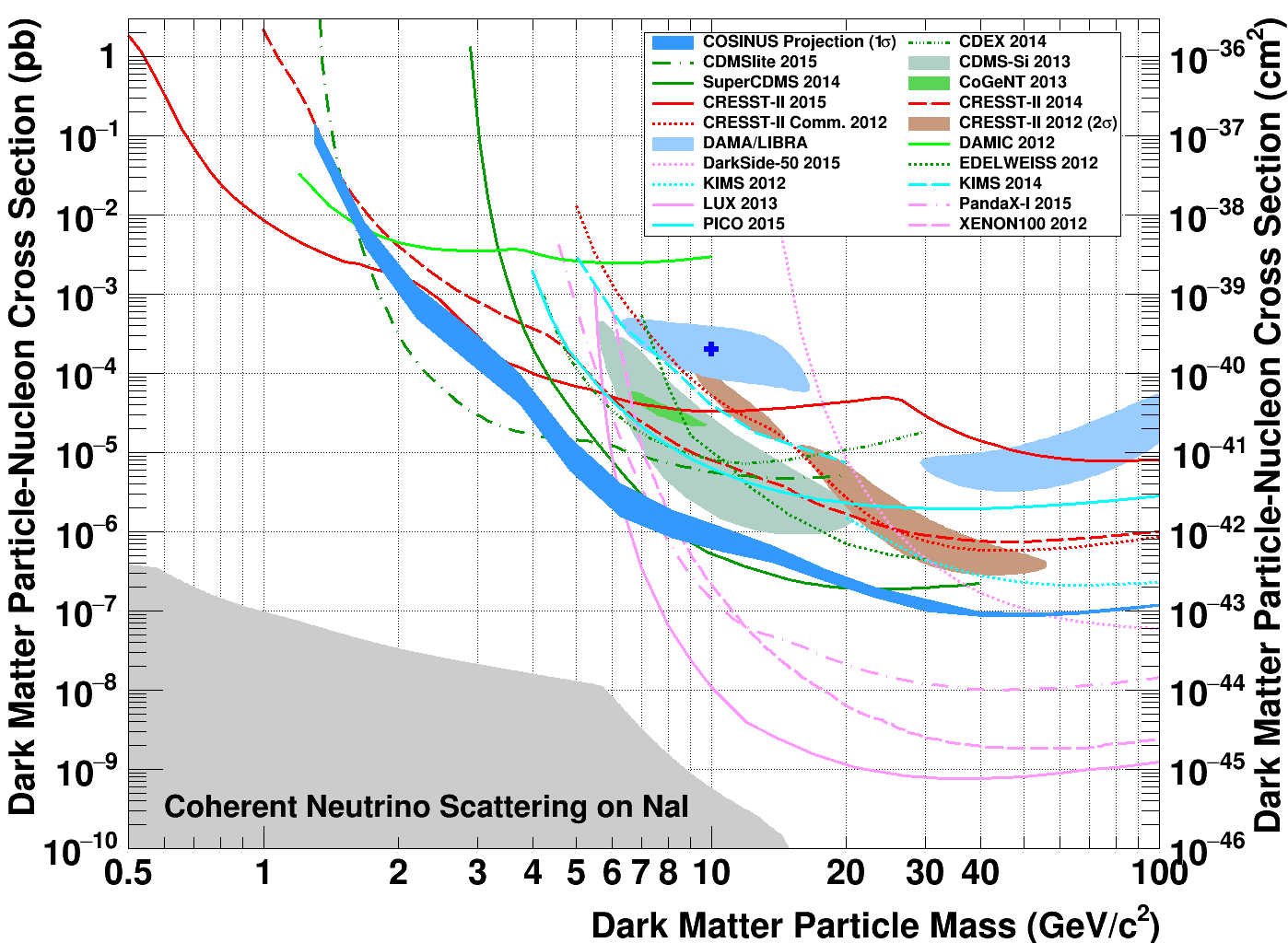}
\caption{Projected sensitivity for a NaI-based experiment using COSINUS detector technology (solid blue band, \unit[1]{$\sigma$} confidence level (C.L.)) for spin-independent elastic dark matter nucleus scattering. Recent results from experiments using silicon and germanium targets are drawn in green \cite{cdex_collaboration_limits_2014,supercdms_collaboration_wimp-search_2015,Agnese2013,SuperCDMS,Aalseth2013,the_damic_collaboration_damic:_2013,edelweiss_collaboration_search_2012}, results from CRESST-II (CaWO$_4$) are depicted in red \cite{CRESST_LISE,CRESST2,brown_extending_2012,CRESST1} and exclusion limits using liquid noble gases in magenta \cite{DarkSide50,Akerib2014,pandax_collaboration_low-mass_2015,Xenon100}. Limits drawn in cyan correspond to bubble chamber technology and experiments with CsI target \cite{pico_collaboration_dark_2015,kims_collaboration_new_2012,lee_search_2014}. The light blue shaded regions correspond to the interpretation of the DAMA/LIBRA (NaI(Tl)) modulation signal by Savage et al. \cite{Savage}. The benchmark point (blue cross) indicates the mass and cross-section chosen for the simulated WIMP contribution presented in \ref{sec:PerformanceSim}. Gray-shaded regions in parameter space will be affected by coherent neutrino nucleus scattering on NaI mainly originating from solar neutrinos \cite{gutlein_impact_2015}.}
\label{fig:landscape}    
\end{figure}
Figure \ref{fig:landscape} depicts the result in the blue colored band (\unit[1]{$\sigma$} confidence level), together with results from other direct dark matter search experiments (see caption and legend). As can be seen, competitive sensitivity to other cryogenic experiments in the range of \unit[(1 to 10)]{GeV/c$^2$} may be reached. To further gain for WIMP masses below that range a significantly reduced threshold would be mandatory, the sensitivity above $\mathcal{O}$(10 GeV/c$^2$) is clearly limited by exposure. \\
The comparison of results from different experiments, as depicted in figure \ref{fig:landscape}, only holds under certain assumptions concerning the dark matter halo and the interaction of dark matter with Standard Model particles. This consideration is further augmented by the use of different target materials, as the impact of the mentioned uncertainties significantly depends on the target material. \\
Obviously, material dependences will be ruled out in the evaluation of COSINUS and DAMA/LIBRA data as both are using a NaI-target. As figure \ref{fig:landscape} clearly shows, the anticipated sensitivity of a COSINUS detector is about two orders of magnitude below the interpretation of the DAMA/LIBRA claim, assuming a standard dark matter halo and elastic WIMP-nucleus scattering (light-blue regions in \ref{fig:landscape} corresponding to recoils off Na and I, respectively \cite{Savage}). \\
The enhanced sensitivity, thereby, is driven by two key factors. Firstly, the particle discrimination via the simultaneous measurement of phonon signal and scintillation. Secondly, the better energy resolution going along with a lower threshold (of \unit[1]{keV}, independent of the type of particle). The latter is of special benefit due to the expected exponential rise of the dark matter recoil spectrum towards low energies. Thus, even with a moderate exposure, COSINUS technology has the potential to add knowledge on the underlying nature of the DAMA/LIBRA signal, in particular on the question whether the signal originates from nuclear recoils or not. \\
For a hypothetical dark matter particle interacting with the electrons of the target material the advantage of particle discrimination vanishes, as the dominant background will also be found in the e$^-$/$\gamma$-band. However, the excellent energy resolution and the low threshold still persist. Thus, a potential future dark matter experiment, increased in target mass and based on COSINUS technology, has promising prospects to give new insight on the long-standing DAMA/LIBRA claim.
\section*{Acknowledgements}
This work was carried out in the frame of the COSINUS R\&D project funded by the  Istituto Nazionale di Fisica Nucleare (INFN) in the Commissione Scientifica Nazionale 5 (CSN5).
\bibliographystyle{h-physrev}
\bibliography{cosinus_1}

\end{document}